\documentclass[final,5p,number,times,twocolumn]{elsarticle}
\usepackage{graphicx}

\newcommand{\Dirac}{\rlap {\hspace{-0.1mm} \slash} D}
\newcommand{\eq}[1]{(\ref{#1})}
\newcommand{\beqn}{\begin{eqnarray}}
\newcommand{\eeqn}{\end{eqnarray}}
\newcommand{\dd}{{\mathrm{d}}}
\newcommand{\cZ}{{\mathcal Z}}
\newcommand{\dD}{{\mathrm{D}}}
\newcommand{\cM}{{\mathcal M}}

\newcommand{\bbbone}{{{\mathchoice {\rm 1\mskip-4mu l} {\rm 1\mskip-4mu l} {\rm 1\mskip-4.5mu l} {\rm 1\mskip-5mu l}}}}

\newcommand{\lr}[1]{ \left( #1 \right) }

\newcommand{\Tr}{ {\rm Tr} \, }
\newcommand{\tr}{ {\rm Tr} \, }

\journal{Nuclear Physics B}

\begin{document}
\sloppy
\begin{frontmatter}

\title{Chiral magnetization of non-Abelian vacuum: a lattice study}

\author[JIPNR,ITEP]{P. V. Buividovich}
\author[LMPT,DMPA,ITEP]{M. N. Chernodub}
\author[ITEP]{E. V. Luschevskaya}
\author[ITEP]{M. I. Polikarpov}
\address[JIPNR]{JIPNR ``Sosny'', National Academy of Science, Acad. Krasin str. 99, Minsk, 220109 Belarus}
\address[ITEP]{Institute for Theoretical and Experimental Physics, B. Cheremushkinskaya 25, Moscow, 117218 Russia}
\address[LMPT]{Laboratoire de Mathematiques et Physique Theorique,
CNRS UMR 6083, F\'ed\'eration Denis Poisson, Universit\'e de Tours, Parc de Grandmont, 37200, France}
\address[DMPA]{Department  of Mathematical Physics and Astronomy, University of Gent, Krijgslaan 281, S9, B-9000 Gent, Belgium}

\date{August 9, 2009}

\begin{abstract}
The chiral magnetization properties of cold and hot vacua are studied using
quenched simulations in lattice Yang-Mills theory. In weak external magnetic
fields the magnetization is proportional to the first power of the magnetic
field. We evaluate numerically the coefficient of the proportionality
(the chiral susceptibility) using near-zero eigenmodes of overlap fermions. We
found that the product of the chiral susceptibility and the chiral condensate
equals to $46(3)\,\mbox{MeV}$. This value is very close to the phenomenological value
of $50\,\mbox{MeV}$. In strong fields the magnetization is a nonlinear function of
the applied magnetic field. We find that the nonlinear features of the
magnetization are well described by an inverse tangent function. The
magnetization is weakly sensitive to temperature in the confinement phase.
\end{abstract}


\end{frontmatter}

\section{Introduction}

Quark is an electrically charged spin-1/2 particle with a magnetic moment.
Following an analogy with electrodynamics~\cite{ref:Landau}, one can conclude
that there exist at least two opposite types of the quark magnetic
back-reaction of the QCD vacuum on the external magnetic field. The
paramagnetic effect enhances magnetic field in the vacuum due to polarization
of the magnetic moments of virtual quarks by the external magnetic field. On
the contrary, the diamagnetic effect weakens the external field due to the
(quantized) transverse orbital motion of the virtual quarks.

Qualitatively, there is a distant similarity of the QCD vacuum to a gas of electrically
charged magnetic dipoles in electrodynamics~\cite{ref:SolidState}. However,
there also is an important difference between QCD and electrodynamics: in the
physically relevant limit of two massless quarks, QCD is classically conformal
theory in which mass scale appears due to nonperturbative quantum effects. The
response of the vacuum on the external magnetic fields is characterized by
dimensionfull quantities and this feature makes the problem essentially
nonperturbative.

In our paper we study, for the first time, the magnetization properties of the
non-Abelian vacuum using numerical simulations in lattice gauge theory. The
problem is not limited by academic interest, because very strong Abelian
magnetic fields can be created, for example, in heavy ion
collisions~\cite{ref:collisions}. Such fields may substantially modify the
phase diagram of QCD, changing even the order of the phase transition from the
hadron phase to the quark-gluon plasma regime~\cite{ref:diagram}. In lattice QCD the
external field methods were already used to evaluate the electric
polarizability of neutral mesons and baryons~\cite{ref:Tiburzi}. The dependence
of the quark condensate on the strength of the uniform magnetic field was also
calculated recently in lattice gauge theory~\cite{ref:chiral:lattice}.

We concentrate on the (para)magnetic response of the vacuum which appears due
to polarization of the spins (magnetic moments) of the virtual quarks and
antiquarks in the external electromagnetic field. A natural quantitative
measure of the spin polarization in the vacuum is given by the expectation
value
\beqn
\langle \bar\Psi \Sigma_{\alpha\beta} \Psi\rangle = \chi(F)\, \langle \bar\Psi
\Psi\rangle\, q F_{\alpha\beta}\,, \label{eq:chi:def}
\eeqn
where
\beqn
\Sigma_{\alpha\beta} = \frac{1}{2 i} [\gamma_\alpha \gamma_\beta - \gamma_\beta
\gamma_\alpha]\,, \label{eq:Sigma}
\eeqn
is the relativistic spin operator, $F_{\alpha\beta}$ is the external electromagnetic field
strength tensor and $q$ is the electric charge of the quark $\Psi$. For
simplicity, we omit flavor indices in \eq{eq:chi:def} and consider it
for one quark flavor.

The quantity~\eq{eq:chi:def} was first introduced by Ioffe and Smilga in Ref.~\cite{ref:IS} in
order to analyze the nucleon magnetic moments which are related to phenomenologically interesting
radiative transitions. Later the value of the magnetic susceptibility was estimated using various
analytical approaches~\cite{ref:others:OPE,ref:Vainshtein,ref:Sasha,ref:Kim}.
The value of the magnetic susceptibility can be measured in experiments on lepton pair photoproduction
through the chiral-odd coupling of a photon with quarks~\cite{ref:phenomenology}, and in
radiative heavy meson decays~\cite{Rohrwild:2007yt}.

The right hand side of Eq.~\eq{eq:chi:def} is proportional to the
electromagnetic field strength tensor $F_{\mu\nu} = \partial_\mu a_\nu -
\partial_\nu a_\mu$ due to the Lorenz covariance. The quark electric charge $q$ appears
in Eq.~\eq{eq:chi:def} since the electromagnetic field $a_\mu$ interacts with
the quark field only in the combination $q a_\mu$. Another proportionality
factor in the right hand side of Eq.~\eq{eq:chi:def} is the chiral condensate
$\langle \bar\Psi \Psi\rangle$ (evaluated at the external electromagnetic field
$F$). This factor allows us to disentangle nonlinear effects of the enhancement
of the chiral condensate in the external magnetic
field~\cite{ref:SushSmi,ref:chiral:lattice} from the effects of the quark's
spin polarization.

In a leading order the magnetization of the QCD vacuum in weak magnetic fields
should be a linear function of the field strength. Using perturbative QCD and the
Schwinger proper time formalism one can show that the magnetization due to a strong
magnetic field at one-loop order is proportional to $B \log B$~\cite{ref:Thomas}.

The strength of the vacuum polarization is characterized by a chiral magnetic
susceptibility $\chi(F)$ in Eq.~\eq{eq:chi:def}. In our discussion we treat the
vacuum magnetization and the quark's spin polarization on equal footing
skipping the $g$--factor which relates these quantities and characterizes the
gyromagnetic ratio of the quarks.

The (chiral) magnetization of the QCD vacuum in the external magnetic
field $B = F_{12} = - F_{21}$ can be described by the dimensionless quantity
\beqn
\mu(q B) = \chi(q B)\, q B\,,
\label{eq:mu}
\eeqn
so that
\beqn
\langle \bar\Psi \Sigma_{12} \Psi\rangle = \mu(q B) \langle \bar\Psi
\Psi\rangle\,, \label{eq:Sigma12}
\eeqn
for the other Lorentz components the polarization~\eq{eq:chi:def} is zero. The
quantity~\eq{eq:mu} is of central interest in our paper.

In Section~\ref{sec:theoretical} we derive an analytical formula which exactly
relates the spinor structure of the low-lying Dirac eigenmodes to the
magnetization~\eq{eq:mu}. This formula is used to evaluate the magnetization
numerically in Section~\ref{sec:numerical}. We discuss in details both the
linear magnetization in a weak field and nonlinear features of this quantity in
stronger fields. Our conclusions are summarized in the last Section.

\section{A magnetization analog of Banks-Casher relation}
\label{sec:theoretical}

In order to calculate the polarization properties of the QCD vacuum we derive
the analytical formula which relates the magnetization to the spin structure of
the low-lying quark eigenmodes. This formula is an analogue of the Banks--Casher
relation~\cite{ref:BanksCasher}
\beqn
\langle \bar\Psi\Psi\rangle = - \lim_{\lambda \to 0} \frac{\pi \rho(\lambda)}{V}\,,
\label{eq:Banks:Casher}
\eeqn
which relates the chiral condensate $\langle \bar\Psi\Psi\rangle$ to the expectation
value of the spectral density of the Dirac eigenmodes $\rho(\lambda)$ (to be defined below).
In Appendix~A we derive the Banks-Casher relation~\eq{eq:Banks:Casher} in order to
illustrate its relation to our analytical result~\eq{eq:chi}.

The Euclidean partition function of QCD is given by the integral
over the gluon fields $A^a_\mu$, $\mu=1,\dots, 4$ and $a=1,\dots, N_c^2-1$, and
over the quark Dirac fields $\Psi_f$, $f=1,\dots N_f$,
\beqn
\cZ_{\mathrm{QCD}} & = & \int \dD A\! \int \dD \Psi\! \int \dD \bar\Psi \,
e^{-S_{\mathrm{YM}}(A) - S_F(A,\Psi,\bar\Psi)} \nonumber\\
& \equiv & \int \dD A\, \det[\Dirac(A) + \cM ]\, e^{-S_{\mathrm{YM}}(A)}\,,
\label{eq:QCD:real}
\eeqn
where $S_{\mathrm{YM}}$ is the Yang--Mills action. The fermion action~is
\beqn
S_F(A,\Psi,\bar\Psi) = \int \dd^4 x \, \bar\Psi_f(x) {[\Dirac(A) + \cM ]}_{ff'}
\Psi_{f'}(x)\,, \label{eq:S:F}
\eeqn
where $\cM$ is the $N_f \times N_f$ mass matrix in the flavor space.

For the sake of simplicity we consider below one fermion species ($N_f = 1$)
with the mass $m$. Then $\cM \equiv m$ and the fermion determinant in
Eq.~\eq{eq:QCD:real} is
\beqn
F(m) = \det[\Dirac(A) + m ] \equiv \prod\limits_k (i \lambda_k(A) + m)\,,
\label{eq:determinant:0}
\eeqn
where $\lambda_k = \lambda_k(A)$ is the eigenvalue of the massless Dirac
operator $\Dirac \equiv \gamma_\mu D_\mu$ in the background of the Euclidean
gauge field configuration $A$. The spectrum of this operator is defined by the
equation
\beqn
\Dirac \psi_k = i \lambda_k \psi_k\,, \label{eq:Dirac:Eq}
\eeqn
where $\psi_k = \psi_k(x;A)$ is the corresponding eigenfunction (below we omit
the argument ``$A$'' in $\psi_k$ and $\lambda_k$ for the sake of simplicity).
Due to the anticommutation property, $\gamma_5 \Dirac + \Dirac \gamma_5 = 0$,
any nonzero eigenvalue, $\lambda_k \neq 0$, comes in a pair with its opposite,
$\lambda_{-k} = - \lambda_k$, corresponding to the eigenfunction $\psi_{-k} =
\gamma_5 \psi_k$.

The eigenfunctions $\psi_k$ form a basis in the spinor space: they are
orthonormalized and complete (we always omit spinor indices),
\beqn
\int \! \dd^4 x \, \psi^\dagger_k(x)\psi_l(x) & = & \delta_{kl}\,,
\label{eq:normalization} \\
\sum_k \psi_k(x) \psi^\dagger_k(x') & = & \delta(x - x')\,.
\label{eq:competeness}
\eeqn

In the thermodynamic limit the expectation value of the
magnetization~\eq{eq:mu} can be expressed via the nonzero eigenmodes,
$\psi_k(x)$, of the massless Dirac operator \eq{eq:Dirac:Eq}.
\beqn
& & - \langle \bar\Psi \Sigma_{\alpha\beta} \Psi\rangle \equiv \langle \Tr
[\Sigma_{\alpha\beta}  \Psi(x) \bar\Psi (x)]\rangle \nonumber
\\
& & = \langle\Tr [\Sigma_{\alpha\beta}  D(x,x)]\rangle = \langle\sum_k
\frac{\psi^\dagger_k(x) \, \Sigma_{\alpha\beta}  \, \psi_k(x)}{i\lambda_k +
m}\rangle
\nonumber \\
& & = \langle\sum_{\lambda_k>0} \frac{\psi^\dagger_k(x) \, \Sigma_{\alpha\beta}
\, \psi_k(x)}{i\lambda_k + m}\rangle
\label{eq:O2} \\
& & \phantom{=} + \langle\sum_{\lambda_k>0} \frac{\psi^\dagger_k(x) \, \gamma_5
\Sigma_{\alpha\beta}  \gamma_5 \, \psi_k(x)}{- i\lambda_k + m}\rangle \,,
\nonumber
\eeqn
where the trace is taken over the spinor indices and the fermion propagator
$D(x,x')$ is defined by Eq.~\eq{eq:propagator}. In the first raw of Eq.~\eq{eq:O2}
the notation $\langle\dots\rangle$ means the average over the gluon, $A_\mu$, and fermion,
$\Psi$, fields. After the integration over the fermion fields is done, the brackets
$\langle\dots\rangle$ mean the average over gluon fields with the weight
$F(m) e^{- S_{\mathrm{YM}}(A)}$, where $F(m)$ is the determinant of the fermion
operator~\eq{eq:determinant:0}. We omit the explicit dependence of the
eigenfunctions $\lambda_k$ and the eigenvalues $\psi_k$ on the gluon field $A_\mu$
and the external electromagnetic field $B$. In the first raw of Eq.~\eq{eq:O2}
the minus sign appears due to anticommutative nature of the fermionic fields.
Following the logic of the derivation of the Banks-Casher formula, we ignore in sums over
eigenvalues the exact zero modes, because the zero modes are inessential in the
thermodynamic limit.

The spin operator~\eq{eq:Sigma} commutes with the $\gamma_5$ matrix,
\beqn
\gamma_5 \Sigma_{\alpha\beta} - \Sigma_{\alpha\beta} \gamma_5 = 0\,,
\eeqn
so that
$\gamma_5 \Sigma_{\alpha\beta} \gamma_5 = \Sigma_{\alpha\beta}$ because $\gamma_5^2 = 1$.
Then Eq.~\eq{eq:O2} gives us the following expression for the magnetization:
\beqn
\langle \bar\Psi \Sigma_{\alpha\beta} \Psi\rangle = 2 m
\langle\sum_{\lambda_k>0} \frac{\psi^\dagger_k(x) \, \Sigma_{\alpha\beta}\,
\psi_k(x)}{\lambda_k^2 + m^2}\rangle\, . \label{eq:lim2}
\eeqn
Similarly to the derivation of the Banks-Casher relation (Appendix~A) we
take the limit $m \to 0$ and get
\beqn
\langle \bar\Psi \Sigma_{\alpha\beta} \Psi\rangle = 2 \pi \langle
\int\limits_0^\infty \dd \lambda \, \nu(\lambda) \delta(\lambda)
\psi^\dagger_\lambda(x) \, \Sigma_{\alpha\beta}\, \psi_\lambda(x)\rangle \,, \
\label{eq:lim3}
\eeqn
where $\nu(\lambda)$ is the spectral density of the Dirac eigenvalues,
\beqn
\rho(\lambda) = \langle \nu (\lambda) \rangle\,, \qquad \nu (\lambda) = \sum_k
\delta(\lambda - \lambda_k)\,.
\label{ref:density:nu}
\eeqn

Then we take the average of this expression over the whole space-time and take the integral over $\lambda$
\beqn
\langle \bar\Psi \Sigma_{\alpha\beta} \Psi\rangle  = \lim_{\lambda \to 0}
\langle\frac{\pi \nu(\lambda)}{V} \int\dd^4 x \, \psi^\dagger_\lambda(x) \,
\Sigma_{\alpha\beta}\, \psi_\lambda(x)\rangle\,. \label{eq:lim4}
\eeqn
Note, that if in Eq.~\eq{eq:lim4} we take the unit operator $\bbbone$ instead
of the spin operator $\Sigma_{\alpha\beta}$ then we immediately recover the
Banks-Casher formula~\eq{eq:Banks:Casher} due to the normalization
condition~\eq{eq:normalization}.

In Appendix B we give some arguments in favor of the validity of the
factorization~\eq{eq:lim4}. Then, using \eq{eq:Banks:Casher}, we get:
\beqn
\langle \bar\Psi \, \Sigma_{\alpha\beta} \, \Psi\rangle = \langle \bar\Psi
\Psi\rangle \langle\int\dd^4 x \, \psi^\dagger_\lambda(x) \,
\Sigma_{\alpha\beta} \, \psi_\lambda(x)\rangle\,.
\label{eq:lim5}
\eeqn
Our sketch of the proof of factorization property~\eq{eq:lim5} in Appendix B is
valid in infinite volume. In the next Section we check numerically the factorization
in our finite volumes as well.

Having compared Eq.~\eq{eq:lim5} with the definition for the magnetic susceptibility $\chi_m$ in
Eq.~\eq{eq:chi:def}, we get
\beqn
\chi(q F) \, q F_{\alpha\beta} = - \lim_{\lambda \to 0} \langle\int\dd^4 x \,
\psi^\dagger_\lambda(x;F) \, \Sigma_{\alpha\beta} \,
\psi_\lambda(x;F)\rangle. \
\label{eq:chi}
\eeqn
Here $\psi_\lambda(x;F)$ is the eigenmode of the Dirac operator in the external (magnetic) background field $B=F_{12}$.
Next, we use the definition~\eq{eq:mu} of the magnetization $\mu$ to rewrite Eq.~\eq{eq:chi} as follows
\beqn
\mu(q B) = - \lim_{\lambda \to 0} \langle\int\dd^4 x \, \psi^\dagger_\lambda(x;B)
\, \Sigma_{12} \, \psi_\lambda(x;B)\rangle\,.
\label{eq:mu:fin}
\eeqn
This is our final analytical expression which we use for the evaluation
of the magnetization in our numerical simulations.

Eq.~\eq{eq:lim5} demonstrates the apparent factorization of the chiral
condensate in line with the original  definition~\eq{eq:chi:def} of
Ref.~\cite{ref:IS}. Due to the factorization~\eq{eq:lim5} the chiral condensate
does not enter explicitly our final formulas~\eq{eq:chi} and \eq{eq:mu:fin},
and therefore one can hope that various ambiguities (related to the definition
and/or logarithmic divergence of the condensate in the quenched limit) does not
enter our definition of the magnetization.

\section{Magnetization from first principles}
\label{sec:numerical}

\subsection{Numerical simulations}

We simulate lattice SU(2) Yang-Mills theory with tadpole-improved Symanzik
action for the gauge fields~\cite{ref:improvedact}. This improvement provides
us with smoother gauge configurations compared to the usual Wilson action.
The fermionic eigenmodes are calculated using the overlap Dirac operator for
quarks in the fundamental representation~\cite{Neuberger:98:1}. One of the most
important advantages of the overlap Dirac operator is its explicit chiral symmetry at all
lattice spacings. Basically, we used the same technique as was implemented in
Ref.~\cite{ref:condensate}.

In our quenched simulations we calculated the magnetization of the $d$-quark
condensate which has the smallest (absolute value of) the electric charge $q =
|q| = e/3$. In continuum notations, the uniform magnetic field in the third
spatial direction is introduced into the Dirac operator by shifting the
non-Abelian vector potential $A_{\mu}$ by the singlet Abelian potential
$a_\mu$:
\begin{eqnarray}
\label{magnetic_field}
A_{\mu}^{ij} & \rightarrow & A_{\mu}^{ij} + a_\mu \delta^{ij}\,, \\
a_{\mu} & = & B/2 \lr{x_{2} \delta_{\mu1} - x_{1} \delta_{\mu2}}.
\nonumber
\end{eqnarray}
In order to adopt the field (\ref{magnetic_field}) to a finite volume with
periodic boundary conditions we have introduced an additional $x$-dependent
boundary twist for fermions following Ref.~\cite{Wiese:08:1}. In a finite
spatial volume $L^3$ with periodic boundary conditions the total magnetic flux
trough any two-dimensional face $L^2$ of the lattice cube should be
quantized~\cite{Damgaard:88:1}. This condition leads to the quantization of the
uniform magnetic field:
\beqn
q B = \frac{2 \pi \, k}{L^{2}}\,,
\qquad
k \in \mathbb{Z}\,.
\label{eq:qB:quant}
\eeqn
The physical strength of the magnetic field is a periodic function of the
flux number $k$ with the period $L^2$. The maximal strength is reached at $k=L^2/2$.

The parameters of our simulations are given in Table~I. For $T=0$ we used three
lattice volumes and two values of the lattice spacing $a$ in order to check the
systematic errors due to finite volume and finite lattice spacing. In our
simulations we used 20 gauge field configurations for each set of parameters
given in Table I.
\begin{table}
\begin{center}
\caption{Parameters of simulations.}
\begin{tabular}{cccccc}
\hline
\hline
\ $L_s$\  & \ $L_s$\  & \ $\beta$\   & \ $a$, fm \  & $V_{3d}$, fm$^3$ & \ $T/T_c$\ \\
\hline
14    & 14   & 3.2810   & 0.103     & $1.44^3$ &  0  \\
16    & 16   & 3.2810   & 0.103     & $1.65^3$ &  0  \\
16    & 16   & 3.3555   & 0.089     & $1.42^3$ &  0  \\
16    &  6   & 3.1600   & 0.128     & $2.01^3$ &  $0.82$  \\
\hline
\hline
\end{tabular}
\end{center}
\label{tbl:param}
\end{table}
We make simulations at zero temperature and at $T=0.82 T_c$. The critical temperature in $SU(2)$
gauge theory is $T_c = 313.(3)$~MeV~\cite{ref:Vitaly}. We evaluate the limit
$\lambda \to 0$ in \eq{eq:mu:fin} by averaging this expression over low-lying
nonzero Dirac eigenmodes with eigenvalues in the interval [0 \dots 50\,MeV].
In order to evaluate the magnetization we use the asymptotic formula~\eq{eq:mu:fin}
which is based on the factorization property~\eq{eq:lim5}. This approach is valid in
the thermodynamic and chiral limits, taken simultaneously.

\begin{figure}[htb]
\begin{center}
  \includegraphics[width=6cm, angle=-90]{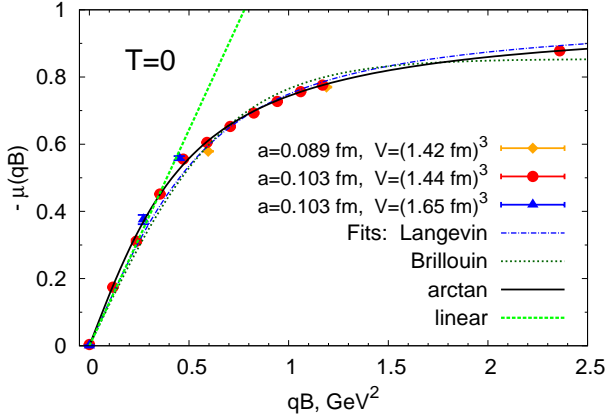}
\end{center}
\vskip -7mm
  \caption{The magnetization $\mu$ as a function of the strength of the external magnetic field $q B$ at zero temperature.
  The lattice spacings and spatial volumes are shown for all data sets. The fits are discussed in the text.}
  \label{fig:m}
\end{figure}

We show the magnetization at zero temperature as the function of the external
magnetic field in Figure~\ref{fig:m}.
The values of the magnetization obtained at different spatial volumes and
different lattice spacings are very close to each other. The relative
discrepancies are much smaller than, e.g., for the values of the chiral
condensate~\cite{ref:chiral:lattice}. This fact indicates finite-volume and
finite-spacing dependencies of bilinear fermionic operators cancel
in~\eq{eq:Sigma12} with a good precision.

\begin{figure}[htb]
\begin{center}
  \includegraphics[width=6cm, angle=-90]{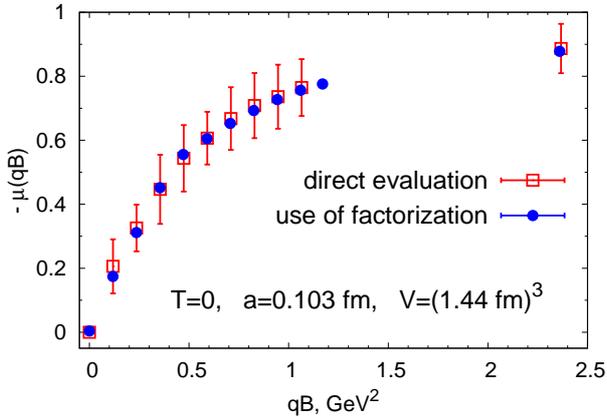}
\end{center}
\vskip -7mm
  \caption{The check of the factorization property for the magnetization $\mu$. The magnetization is calculated
  using both the original~\eq{eq:lim2} and the factorized~\eq{eq:mu:fin} definitions (shown as the empty squares
  and the full circles, respectively).}
  \label{fig:f}
\end{figure}

In Figure~\ref{fig:f} we have also checked that the factorization property~\eq{eq:lim5}
is established very well for all checked strengths of the magnetic field. The values of the magnetization,
evaluated with the help of the original~\eq{eq:lim2} and factorized~\eq{eq:mu:fin}
definitions agree with each other within error bars. Unfortunately,
the nonfactorized definition~\eq{eq:lim2} gives us much larger error bars compared
to the factorized definition~\eq{eq:mu:fin} at the same statistics. Therefore below we
use the factorized definition only.

The behavior of the magnetization is consistent with general expectations: at
low magnetic fields the magnetization is linear. This fact indicates the existence
of a nonzero susceptibility at vanishingly small external magnetic field confirming
the presence of the paramagnetic contribution due to the quark's spin. At high
magnetic fields the quarks ensembles should be fully polarized so that the
magnetization should come to a saturation regime. Mathematically, in Eq.~\eq{eq:mu:fin}
at strong magnetic fields the lowest-lying eigenfunctions with $m=+1$ eigenvalue of
the spin projection operator $\Sigma_{12}$ become dominated over the eigenfunctions with
the $m=-1$ eigenvalue, thus leading to the full polarization of the eigenmodes. In our
units the saturation condition is to be as follows:
\beqn
\lim_{q B \to \infty} \mu(q B) = - 1\,.
\label{eq:high:qH}
\eeqn
Figure~\ref{fig:m} supports this observation.

\begin{figure}[htb]
\begin{center}
  \includegraphics[width=6cm, angle=-90]{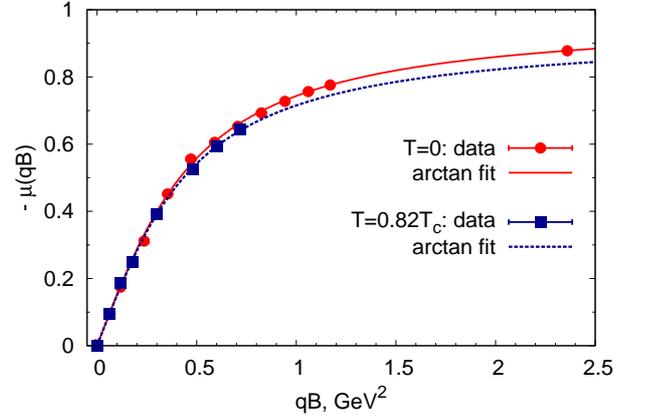}
\end{center}
\vskip -7mm
  \caption{The magnetization $\mu$ as a function of the strength of the magnetic field, $q B$ at
  $T=0$ ($14^4$ lattice) and $T=0.82 T_c$  ($16^3 \times 6$ lattice).
  Best arctan-based fits~\eq{eq:arctan2} with two fitting parameters are shown.}
  \label{fig:mt}
\end{figure}

In Figure~\ref{fig:mt} we compare our zero-temperature data with the finite temperature magnetization obtained at $T=0.82 T_c$.
Visually, the effect of the temperature on magnetization is rather small. More
detailed analysis of $T=0$ and $T>0$ data will be done in the next subsection.

\subsection{Analysis of magnetization}

Let us compare our results with well known expressions for simple paramagnetic systems.
The specific (i.e., per atom) magnetization of a classical ideal paramagnetic gas
is given by~\cite{ref:SolidState}
\beqn
\mu^{\mathrm{class}}(B) & = & \mu \left[\coth \alpha(B) - \frac{1}{\alpha(B)} \right]\,,
\label{eq:langevin}
\\
\alpha(B) & = & \frac{\mu B}{k_B T}\,,
\label{eq:alpha}
\eeqn
where $\mu$ is the magnetic moment of a single molecule, $T$ is temperature and $k_B$ is the Boltzmann constant.
The function in the brackets in Eq.~\eq{eq:langevin} is often called the Langevin function.

The quantum analogue of the Langevin function~\eq{eq:langevin} for a spin 1/2-particle is described
by the Brillouin behavior~\cite{footnote1}:
\beqn
\mu^{\mathrm{quant}}(B) = \mu \left[2 \coth 2 \alpha(B) - \coth \alpha(B)\right]\,.
\label{eq:brillouin}
\eeqn
The atoms are supposed to be electrically neutral otherwise the paramagnetic magnetization -- either classical~\eq{eq:langevin}
or quantum~\eq{eq:brillouin} -- should be supplemented by a diamagnetic contribution due to Landau quantization (the motion of the
atoms in transverse direction to the magnetic field is to be constrained to the Landau levels). In an electron gas the diamagnetic
contribution provides an essential part of the total magnetization~\cite{ref:Landau}. However, we do not consider the diamagnetic
contribution of the quark's motion because of the paramagnetic nature of the condensate~\eq{eq:chi:def}.

A similar behavior of the magnetization \eq{eq:langevin} and \eq{eq:brillouin}
is also provided by inverse tangent function which we add here for completeness:
\beqn
\mu^{\mathrm{trig}}(B) = \frac{2\mu}{\pi}  \arctan \alpha(B)\,.
\label{eq:arctan}
\eeqn

At weak magnetic fields all functions \eq{eq:langevin}, \eq{eq:brillouin} and \eq{eq:arctan} are linear in the magnetic field.
At strong magnetic field all expressions for the magnetization are consistent with the saturation
property~\eq{eq:high:qH}.

In the examples \eq{eq:langevin}, \eq{eq:brillouin} and \eq{eq:arctan} the temperature $T$
enters the magnetization via Eq.~\eq{eq:alpha}. The temperature plays a role of the disorder
ingredient which diminishes the magnetization of the system. In our simulations the disorder
factor is coming -- even at the zero-temperature case -- from the random fluctuating
background of the non-Abelian gauge fields, making the temperature $T$ to be an effective
parameter in all considered examples.

We fit our data by three functions inspired by Eqs.~\eq{eq:arctan}, \eq{eq:langevin} and \eq{eq:brillouin}:
\beqn
\mu^{\mathrm{class}}_{\mathrm{fit}}(B) & = & \mu_\infty \left[\coth \frac{3 \chi_0 q B}{\mu_\infty} - \frac{\mu_\infty}{3 \chi_0 q B} \right]\,,
\label{eq:langevin2}\\
\mu^{\mathrm{quant}}_{\mathrm{fit}}(B) & = & \mu_\infty \left[2 \coth \frac{2 \chi_0 q B}{\mu_\infty} - \coth \frac{\chi_0 q B}{\mu_\infty}\right]\,, \quad  \label{eq:brillouin2}\\
\mu^{\mathrm{trig}}_{\mathrm{fit}}(B) & = & \frac{2 \mu_\infty}{\pi}  \arctan \frac{\pi \chi_0 q B}{2 \mu_\infty}\,.
\label{eq:arctan2}
\eeqn
These fitting functions share the following properties. All these fitting functions are the functions of the two parameters:
the zero-field susceptibility $\chi_0$ and the strong-field saturation constant, $\mu_\infty$.
Indeed, in weak and strong field limits one gets, respectively
\beqn
\mu^i_{\mathrm{fit}}(B) = \chi_0 \cdot q B + \dots \qquad \mbox{as} \quad q B \to 0\,;
\label{eq:linear}
\eeqn
and~\cite{footnote2}
\beqn
\lim_{qB \to \infty} \mu^i_{\mathrm{fit}}(q B) = - \mu_\infty\,.
\label{eq:limit:qB}
\eeqn
Here the index $i$ stands for the type of the fitting function: ``class'', ``quant'' or ``trig''.

According to Eq.\eq{eq:high:qH} we expect that
\beqn
\mu_\infty = 1\,. \label{eq:mu:exp}
\eeqn
Thus we fit the data using the functions \eq{eq:langevin2}, \eq{eq:brillouin2}
and \eq{eq:arctan2} with $\chi_0$ and $\mu_\infty$ being the two fitting
parameters. We also perform an additional set of one-parameter fits with fixed $\mu_\infty = 1$.
Finally, we also fit the weak-field behavior of the magnetization by the linear
function~\eq{eq:linear}.

The fitting results are presented in Tables~\ref{tbl:results}
and \ref{tbl:results:t} for zero and non-zero temperature cases, respectively.
\begin{table}[htb]
\begin{center}
\caption{Fitting results at $T=0$.}
\label{tbl:results}
\begin{tabular}{lccc}
\hline
\hline
fits        & $-\chi_0$, GeV${}^{-2}$ & $\mu_\infty$ & $\chi^2/$d.o.f. \\
\hline
\multicolumn{4}{c}{one parameter fits} \\
linear      & 1.42(6) & - & 25  \\
arctan      & 1.43(1) & 1 & 285 \\
Langevin    & 1.22(2) & 1 & $10^4$   \\
Brillouin   & 0.71(6) & 1 & $6\cdot 10^5$  \\
\hline
\multicolumn{4}{c}{two parameter fits} \\
arctan      & \bf{1.547(6)} & 0.985(7) & 7.2  \\
Langevin    & 1.37(2)  & 0.976(2) & 73  \\
Brillouin   & 1.06(3)  & 0.883(4) & $10^3$  \\
\hline
\hline
\end{tabular}
\end{center}
\end{table}
\begin{table}[htb]
\begin{center}
\caption{Fitting results at $T = 0.82 T_c$.}
\label{tbl:results:t}
\begin{tabular}{lccl}
\hline
\hline
fits        & $-\chi_0$, GeV${}^{-2}$ & $\mu_\infty$ & $\chi^2/$d.o.f. \\
\hline
\multicolumn{4}{c}{one parameter fits} \\
linear      & 1.46(6) & - & 1.8  \\
arctan      & 1.46(2) & 1 & 1.05  \\
Langevin    & 1.36(4) & 1 & 3.8  \\
Brillouin   & 1.19(6) & 1 & 17.6  \\
\hline
\multicolumn{4}{c}{two parameter fits} \\
arctan      & \bf{1.53(3)} & 0.94(2) & 0.57  \\
Langevin    & 1.49(4) & 0.87(2) & 0.71  \\
Brillouin   & 1.47(4) & 0.71(2) & 1.1  \\
\hline
\hline
\end{tabular}
\end{center}
\end{table}

At $T=0$ all fits have quite large values of $\chi^2/$d.o.f., Table~\ref{tbl:results}.
The lowest value of this important quantity is reached for the arctan fitting function~\eq{eq:arctan2}
with two fitting parameters. Notice that both for the arctan-based~\eq{eq:arctan2} function and for the
classical Langevin~\eq{eq:langevin2} fitting function the asymptotic values of the polarization $\mu_\infty$
are very close to the theoretical expectation~\eq{eq:mu:exp}.

At $T=0$ the quantum (Brillouin) function~\eq{eq:brillouin2}
does not work at all: it has the very large value of $\chi^2/$d.o.f. and the strong-field limit is also inconsistent with
our expectation~\eq{eq:mu:exp}. The fitting functions -- corresponding to the one-parameter fits -- are shown in
Figure~\ref{fig:m} by lines.

We point out that the linear fit at the weak field limit does not agree with most other
fitting functions. The linear fitting is done for the relatively weak magnetic fields,
$0 \leqslant \sqrt{q H} \leqslant 500 \, \mbox{MeV}$ (shown as a straight line in Figure~\ref{fig:m}).
The fields from this interval, however, are of the order or even
higher than typical QCD scale $\Lambda_{\mathrm{QCD}} \sim 200$\,MeV. Therefore, the nonlinear effects may
affect the determination of the zero-field susceptibility if the fitting is done by the linear function.
Below we quote the result for the magnetic susceptibility using the arctan-based~\eq{eq:arctan2} function.
The corresponding value is presented in the bold font in Table~\ref{tbl:results}.

In a finite volume the physical magnetic field~\eq{eq:qB:quant}
is a periodic function of the number of elementary magnetic fluxes $k$ going through any face of the lattice
volume~\cite{Damgaard:88:1}. Since the periodic finite-volume behavior at very strong fields is not reflected
in the form of the fitting functions~\eq{eq:langevin2}, \eq{eq:brillouin2} and \eq{eq:arctan2}
the asymptotic magnetization $\mu_\infty$ of two-parameter fits may deviate from the expected high-field limit~\eq{eq:mu:exp}.
However, the artifacts related to the finite volume of the lattice are small since for the best fits by the
functions~\eq{eq:langevin2} and \eq{eq:arctan2} the deviations from the limit~\eq{eq:mu:exp} are small.
We also note that the logarithmic effects -- predicted in Ref.~\cite{ref:Thomas} -- cannot be reliably determined
from our data because of the coarse grid of the data points at strong fields.

At $T=0.82 T_c$ the fits have more reasonable values of $\chi^2/$d.o.f., Table~\ref{tbl:results:t}.
The error bars of the best fit parameters are larger so that the zero-field susceptibility agrees
within error bars for almost all the fitting functions except for the Brillouin function~\eq{eq:brillouin2}.
The best fit -- in terms of both the quality of the fit and the value of the asymptotic polarization
$\mu_\infty$, Eq.~\eq{eq:mu:exp} -- is the arctan-based~\eq{eq:arctan2} function. The accepted value is given
in the bold font in Table~\ref{tbl:results:t}. Finally, the linear fitting is done in the
range $0 \leqslant \sqrt{q H} \leqslant 0.425 \mbox{MeV}$.

The best two parameter arctan-fits for both values of temperature are shown in Figure~\ref{fig:mt}
by lines along with the data. The magnetic susceptibility turns out to be insensitive with respect to
the variation of the temperature:
\beqn
\chi_0 =
\left\{
\begin{array}{lll}
- 1.547(6) \!\!& \mbox{GeV}^{-2} & \quad T=0 \\
- 1.53(3)  \!\!& \mbox{GeV}^{-2} & \quad T=0.82 T_c
\end{array}
\right.
\label{eq:summary}
\eeqn
Our lattice spacings, $a = \Lambda_{\mathrm{UV}}^{-1}$, correspond to the scales $\Lambda_{\mathrm{UV}}(T=0) \sim 2\,\mbox{GeV}$
and $\Lambda_{\mathrm{UV}}(T=0.82 \,T_c) \sim 1.5\,\mbox{GeV}$, respectively. Other estimations of the chiral susceptibility
were done in~\cite{ref:IS,ref:others:OPE,ref:Vainshtein,ref:Sasha,ref:Kim,Rohrwild:2007yt,Ioffe:2009yi}.

Theoretically, the value of the magnetic susceptibility can be parameterized in the form~\cite{ref:Vainshtein}
\beqn
\chi = - \frac{c_\chi N_c}{8 \pi^{2} f_\pi^2}\,,
\label{eq:chi:theory}
\eeqn
were $c_\chi$ is a dimensionless parameter and $f_\pi=130.7$\,MeV is the pion
decay constant for $N_c=3$.

In the notations of Eq.~\eq{eq:chi:theory} the result of our calculation~\eq{eq:summary} at $T=0$ corresponds to
\beqn
c_\chi(T=0)=1.043(4)\,.
\label{eq:c}
\eeqn
The operator product expansion combined with the pion dominance idea gives us the value $c_\chi=2$~\cite{ref:Vainshtein}.
The corresponding theoretical prediction for $SU(2)$ gauge theory is
\beqn
\chi(T=0) = - 2.97\,{\mbox{GeV}}^{-2} \qquad (\mbox{for}\ N_c=2)\,.
\label{eq:theory}
\eeqn
The holographic description of QCD gives a slightly higher value for the susceptibility, $c_\chi=2.15$ \cite{ref:Sasha}.
Both the results of Refs.~\cite{ref:Vainshtein} and \cite{ref:Sasha} agree well with the original QCD sum rule fit
made by Ioffe and Smilga~\cite{ref:IS}.

In the instanton liquid model of the QCD vacuum the magnetic susceptibility was first calculated in Ref.~\cite{ref:Kim}.
After a proper rescaling we obtain the value $c_\chi=1.24$ corresponding to the instanton model.

Thus, our zero-temperature result~\eq{eq:summary}, \eq{eq:c} is by $25\%$ smaller than the value of the magnetic susceptibility
obtained in the instanton vacuum~\cite{ref:Kim}, and is by a factor of two smaller
than the value obtained by traditional field theoretic and modern holographic approaches.
These discrepancies are not unexpected since we used the quenched lattice study in which all vacuum quark loops are ignored,
and the anomalous dimension of the chiral susceptibility was not taken into account. Moreover, our calculations are performed in
the $SU(2)$ gauge theory in which the number of colors is reduced in comparison with the real QCD.

An experimentally relevant and phenomenologically interesting quantity is given by the product of the chiral
susceptibility $\chi$ and the chiral condensate $\langle \bar\Psi\Psi\rangle$~\cite{ref:phenomenology}. Using
our zero-temperature result for the chiral susceptibility~\eq{eq:summary} and the value for the chiral condensate obtained
in other quenched studies~\cite{ref:chiral:lattice,ref:condensate}, $\langle \bar\Psi\Psi\rangle = {[310(6) \, \mbox{MeV}]}^3$,
one gets
\beqn
-\chi \, \langle \bar\Psi\Psi\rangle = 46(3)\,\mbox{MeV}\ \qquad \mbox{[quenched limit]}\,.
\label{eq:chi:cond}
\eeqn
This result is surprisingly close to the estimation based on the QCD sum rules techniques,
which gives for \eq{eq:chi:cond} the number of the order of 50~MeV~\cite{ref:others:OPE}.

\section{Conclusions}

We have evaluated for the first time the magnetic susceptibility of the chiral condensate using the first-principle
methods of lattice $SU(2)$ gauge theory in the quenched limit. To this end we have derived formula~\eq{eq:chi} which
relates the chiral magnetization of the QCD vacuum to the low-lying chiral eigenmodes of the Dirac operator in
a manner of the Banks-Casher relation (the exact zero modes do not contribute to the magnetization in the thermodynamic
limit). In order to derive Eq.~\eq{eq:chi} we used the factorization property~\eq{eq:lim5}, which was verified numerically.
We calculated these eigenmodes using the overlap fermion operator in the background of gluon fields generated
with the help of the tadpole-improved Symanzik action.

We found that at weak magnetic field the magnetization of the QCD vacuum is a linear function of
the field strength. The associated chiral magnetic susceptibility is almost
independent on temperature~\eq{eq:summary} up to $T = 0.82\, T_c$. The value of the
magnetization~\eq{eq:summary} is smaller compared to the existing
analytical estimates~\eq{eq:theory} for $SU(2)$ gauge theory. We attribute the
reason for the difference to the quenching effects.

The nonlinear features of the magnetization are very well described by an inverse tangent function~\eq{eq:arctan2} of
the applied magnetic field, Figure~\ref{fig:mt}. This parametrization of the magnetization works
both at zero and non-zero temperatures in the confinement phase.

\section*{Acknowledgments}

The authors are grateful to Ph.~Boucaud, V.G.~Bornyakov, V.V.~Braguta,
A.S.~Gorsky, B.L.~Ioffe, B.O.~Kerbikov, D.~Kharzeev, A.~Krikun, S.M.~Morozov,
V.A.~Novikov, V.I.~Shevchenko, M.I.~Vysotsky, and V.I.~Zakharov for interesting
discussions. We thank B.~Pire for making us aware of
Ref.~\cite{ref:phenomenology}. This work was partly supported by Grants RFBR
No. 08-02-00661-a, grant for scientific schools No. NSh-679.2008.2, by the
Russian Federal  Agency for Nuclear Power, and by the STINT Institutional grant
IG2004-2 025. P.V.B. is also partially supported by the Euler scholarship from
DAAD, by a scholarship of the Dynasty Foundation and by the grant BRFBR
F08D-005 of the Belarusian Foundation for Fundamental Research. E.V.L. is
partially supported by grant for scientific schools No Nsh-4961.2008.2 and
grants  RFBR 06-02-17012 and 09-02-00629. The calculations were partially done
on the MVS 50K at Moscow Joint Supercomputer Center.

\appendix

\section{Banks-Casher relation: chiral condensate via eigenmodes}

In Section~\ref{sec:theoretical} the relation of the magnetization to the Dirac
eigenmodes was shown to be given by Eq.~\eq{eq:chi}. For the sake of completeness
we present below a derivation of the Banks-Casher relation~\cite{ref:BanksCasher} which is
very similar to the derivation of Eq.~\eq{eq:chi}.

A differentiation of the partition function~\eq{eq:QCD:real} with respect to
the mass $m$ in the thermodynamic limit provides us with the chiral condensate,
\beqn
\langle \bar\Psi\Psi\rangle = - \frac{1}{V} \frac{\partial}{\partial m} \ln
\cZ_{\mathrm QCD} = - \frac{1}{V} \Bigl\langle\sum\limits_{\lambda_k > 0}
\frac{2 m}{\lambda_k^2 + m^2} \Bigr\rangle\,, \label{eq:condensate:m}
\eeqn
where we have used Eq.~\eq{eq:determinant:0} as well as the pairwise appearance
of the mutually opposite eigenvalues,
\beqn
\frac{1}{F(m)} \frac{\partial F(m)}{\partial m} = \sum\limits_{\lambda_k}
\frac{1}{i \lambda_k + m} = \sum\limits_{\lambda_k > 0} \frac{2 m}{\lambda_k^2
+ m^2}\,. \label{eq:determinant:diff}
\eeqn
The zero mode(s), corresponding to $\lambda_0 = 0$, are not counted in
Eq.~\eq{eq:determinant:diff} because they give vanishing contribution in the
thermodynamic limit, and thus are irrelevant.

In the chiral limit, $m \to 0$, Eq.~\eq{eq:condensate:m} provides us with the
celebrated Banks-Casher formula~\cite{ref:BanksCasher}, Eq.~\eq{eq:Banks:Casher},
which relates the chiral condensate $\langle \bar\Psi\Psi\rangle$ to
the expectation value $\rho(\lambda)$ of the spectral density $\nu(\lambda)$
of the Dirac eigenvalues~\eq{ref:density:nu} in the limit to zero
virtuality, $\lambda \to 0$. In order to derive Eq.~\eq{eq:Banks:Casher} we
have used the relations
\beqn
\lim_{m \to 0} \frac{1}{\pi} \frac{m}{\lambda^2 + m^2} = \delta(\lambda) \quad
\mbox{and} \quad \int\limits_0^\infty \dd \lambda \, \delta(\lambda) =
\frac{1}{2}\,, \label{eq:useful:delta}
\eeqn
so that
\beqn
& & \hspace{-1cm} \sum\limits_{\lambda_k > 0} \frac{2 m}{\lambda_k^2 + m^2} =
\int\limits_0^\infty \dd \lambda \, \nu(\lambda)\,
\frac{2 m}{\lambda^2 + m^2}  \nonumber\\
& & \mathop{\overrightarrow{\hskip 10mm}}_{{m \to 0}} 2 \pi
\int\limits_0^\infty \dd \lambda \, \nu(\lambda) \delta(\lambda) = \pi
\lim_{\lambda \to 0} \nu(\lambda)\,. \label{eq:m0}
\eeqn

The fermionic propagator in the background of the gauge field $A_\mu$
can be expressed in terms of the eigenmodes~\eq{eq:Dirac:Eq} as follows:
\beqn
\langle \Psi(x) \bar\Psi(x') \rangle_{A} \equiv D(x,x') = \sum_k
\frac{\psi_k(x) \psi^\dagger_k(x')}{i\lambda_k + m}\,. \label{eq:propagator}
\eeqn
Due to the completeness condition~\eq{eq:competeness}
the propagator~\eq{eq:propagator} satisfies the equation
\beqn
[\Dirac(A) + m] \, D(x,x') = \delta(x-x')\,. \label{eq:propagatoreq}
\eeqn

\section{Factorization in Eq.~\eq{eq:lim5}}

Below discuss a possible origin of the factorization in Eqs. \eq{eq:lim4} and
\eq{eq:lim5}, which was used to evaluate the chiral susceptibility in this article.
Consider two local operators, ${\cal O}_1(x)$ and ${\cal O}_2(x)$,
which are functions of the lattice coordinate~$x$. Then in infinite volume limit
\begin{equation}
\Bigl\langle \frac 1V \sum_{x} {\cal O}_1 (x)\, {\cal O}_2(y) \Bigr\rangle
= \langle {\cal O}_1 \rangle \, \langle {\cal O}_2 \rangle \,.
\label{eq:AB}
\end{equation}
The proof of \eq{eq:AB} is trivial, since the quantum average in lattice calculations is equivalent
to a sum over infinite number of gauge field configurations:
\beqn
& & \Bigl\langle \frac{1}{V} \sum_{x} {\cal O}_1(x)\, {\cal O}_2(y) \Bigr\rangle
\nonumber \\
& & \hskip 5mm =
\lim_{N_{\mathrm{conf}}\to\infty} \sum_{i=1}^{N_{\mathrm{conf}}} \frac{1}{V} \sum_{x} {\cal O}_1(x)\, {\cal O}_2 (y)\, ,
\label{eq:sumconf}
\eeqn
where $V$ is the number of lattice points $x$, and the sum in \eq{eq:sumconf}
goes over gauge field configurations labeled by the index $i$.
The gauge field configurations are generated with the Boltzmann probability
density $e^{-S_{YM}(A)}/{\cal{Z}}$. For infinite volume, $V \to \infty$, the
average for one gauge field configuration is equal to the quantum average,
\beqn
\frac{1}{V} \sum_{x} {\cal O}_1(x) = \langle {\cal O}_1(x) \rangle \, . \label{eq:average}
\eeqn
Equation~\eq{eq:AB} follows from Eqs.~\eq{eq:sumconf} and \eq{eq:average}.

If one takes ${\cal O}_1 (x)={\cal O}_2 (x)=\tr F_{\mu\nu}^2(x)$, then in the continuum limit
we get from Eq.~\eq{eq:AB} the well known factorization formula:
\beqn
\lim_{V \to \infty} \Bigl\langle \frac{1}{V} \int d^4 x\, \tr F_{\mu\nu}^2(x)\, \tr F_{\mu\nu}^2(y) \Bigr\rangle
= \langle \tr F_{\mu\nu}^2\rangle^2\,. \
\eeqn
The factorization of the magnetization, Eqs.~\eq{eq:lim4} and \eq{eq:lim5}, is very
similar to the factorization~\eq{eq:AB}. We only have to prove that the analogue of
Eq.~\eq{eq:average} is valid for the bulk quantities $\lim_{\lambda \to 0} \pi \nu(\lambda)/V $ and
$\lim_{\lambda \to 0}\int\dd^4 x \, \psi^\dagger_\lambda(x) \,
\Sigma_{\alpha\beta}\, \psi_\lambda(x)$. One can naturally assume that these quantities have a smooth
behavior towards the continuum limit, $V\to \infty$. Then the average over one infinitely large
configuration of the gauge fields is equal to the usual quantum average of the factorized expression, and then the factorization~\eq{eq:lim5} is valid.


\begin{thebibliography}{99}
\bibitem{footnote1}
Here $\mu$ and $\alpha$ have different normalizations compared to the ones used in Eq.~\eq{eq:langevin}.
This remark plays no role in our discussion below since we are interested only in the general functional
behavior of the magnetization.
\bibitem{footnote2}
The minus sign appears in the right hand side of Eq.~\eq{eq:limit:qB} because the weak-field susceptibility
is negative, $\chi_0 < 0$.
\end{thebibliography}
\end{document}